\documentclass[12pt,preprint]{aastex}

\usepackage{lineno}

\usepackage{epsfig}

\usepackage{xcolor}

\usepackage{amsmath,amssymb,gensymb}

\usepackage{soul}

\usepackage{ifsym}
\usepackage{wasysym}
\usepackage{MnSymbol}

\begin{document}

%
%
%

\title{Testing  Einstein's Equivalence Principle and its Cosmological Evolution from Quasar Gravitational Redshifts}

%

\author{E. MEDIAVILLA\altaffilmark{1,2} \& J. JIM\'ENEZ-VICENTE\altaffilmark{3,4}}
%

%

\altaffiltext{1}{Instituto de Astrof\'{\i}sica de Canarias, V\'{\i}a L\'actea S/N, La Laguna 38200, Tenerife, Spain}
\altaffiltext{2}{Departamento de Astrof\'{\i}sica, Universidad de la Laguna, La Laguna 38200, Tenerife, Spain}
\altaffiltext{3}{Departamento de F\'{\i}sica Te\'orica y del Cosmos, Universidad de Granada, Campus de Fuentenueva, 18071 Granada, Spain}
\altaffiltext{4}{Instituto Carlos I de F\'{\i}sica Te\'orica y Computacional, Universidad de Granada, 18071 Granada, Spain}
%

\begin{abstract}
We propose and apply a new test of Einstein's Equivalence Principle (EEP) based on the gravitational redshift induced by the central super massive black hole of quasars in the surrounding accretion disk. Specifically, we compare the observed gravitational redshift of the Fe III$\lambda\lambda$2039-2113 emission line blend in quasars with the predicted values in a wide, uncharted, cosmic territory  ($0 \lesssim z_{cosm}\lesssim3$).  For the first time we measure,  with statistical uncertainties comparable or better than those of other classical methods outside the Solar System, the ratio between the observed  gravitational redshifts and the theoretical predictions in 10 independent cosmological redshift bins in the  $1 \lesssim z_{cosm}\lesssim3$ range. {The average of the measured over predicted gravitational redshifts ratio in this cosmological redshift interval is $\langle z^m_g/z_g^p\rangle=1.05\pm 0.06$ with scatter $0.13\pm 0.05$ showing no cosmological evolution of EEP within these limits. This method can benefit from  larger samples of measurements with better S/N ratios, paving the way for high precision tests (below 1\%) of EEP on cosmological scales.}

\end{abstract}


\section{Introduction \label{intro}}

Einstein's Equivalence Principle (EEP) is the central premise of gravitation. It meshes gravity with all the laws of physics and, in particular, predicts the gravitational redshift of photons. EEP is very well tested via gravitational redshift experiments in the solar neighborhood (Pound \& Snider 1964, Vessot et al. 1980, Roca Cort{\'e}s \& Pall{\'e} 2014, Joyce et al. 2018, Gravity Collaboration et al. 2018, Gonz\'alez Hern\'andez et al. 2020), but {very scarcely on cosmic scales}{.  Departures from EEP are not usually allowed in cosmological tests of gravitation (Bonvin \& Fleury 2018), and when taken into consideration ({Rapetti et al. 2010, Reyes et al. 2010}), the results are not independent of the assumption of the $\rm \Lambda$CDM model (Wojtak et al. 2011).}

A notable exception to the lack of cosmological tests of EEP based on astrophysical phenomena {(albeit  limited to a very narrow range of cosmological redshift, $z_{cosm}\simeq0.22$)} is the study of the gravitational redshift effect in clusters of galaxies (Wojtak et al. 2011, see also Dom\'\i nguez-Romero et al. 2012, Kaiser 2013, Jimeno et al. 2015, Sadeh, Feng \& Lahau 2015, Alam et al. 2017). The basis of this test is to average the emission line shifts of a huge number of cluster member galaxies moving in the gravitational potentials of their clusters to detect very small gravitational redshift differences ($\Delta \sim -10\rm\,km\,s^{-1}$) between the outskirts and the centers of the clusters. Wojtak et al. (2011) obtained an integrated value of $z^m_g/ z_g^p\sim 0.9\pm0.3$ for the ratio between the measured and predicted gravitational redshifts in agreement with EEP. Subsequent studies have shown, however, that this measurement is affected by transverse Doppler (Zhao et al. 2013), an effect that can have the same order of magnitude than the gravitational redshift for systems in virial equilibrium, and by light cone and surface brightness effects (Kaiser 2013, Jimeno et al. 2015, Sadeh, Feng \& Lahau 2015).  Taking into account these effects, the initial prediction of Wojtak et al. (2011) modifies to ${z^m_g/ z_g^p}\sim 0.6\pm0.3$
(Kaiser 2013). Sadeh, Feng \& Lahau (2015) follow up the analysis by Wojtak et al. (2011) with a larger dataset  finding an average redshift of $\Delta = -11^{+7}_{-5} \rm\,km\,s^{-1}$, consistent within uncertainties with Wojtak et al. (2011) and in good agreement with the prediction ($\Delta \sim -12 \rm\,km\,s^{-1}$) by Kaiser et al. (2013), but with larger uncertainties  ($z^m_g/ z_g^p\sim 0.9^{+0.6}_{-0.4}$). Jimeno et al. (2015) examine three major cluster samples from the SDSS finding statistical agreement with the model predictions for two of them ($z^m_g/ z_g^p\sim 0.95\pm0.3$ for GMBCG and $z^m_g/ z_g^p\sim 0.7\pm0.2$ for redMaPPer) but anomalous measurements ($\Delta \sim +2.5 \rm\,km\,s^{-1}$ measured vs. $\Delta \sim -20 \rm\,km\,s^{-1}$ predicted) for WHL12. 

%

We propose here a different astrophysical scenario where the gravitational redshift effect is stronger by up to three orders of magnitude: the accretion disks formed by ionized gas rotating around the super massive black holes (SMBH) located at the centers of active galactic nuclei (AGN) and quasars. For typical distances from the SMBH  of $\sim10$ light-days and masses of the SMBH of  $\sim 10^9M_\odot$ we expect gravitational redshifts $z_g \sim GM/c^2R\sim 0.006$ equivalent to $\sim 1700\rm\,km\,s^{-1}$. Recently, we have discovered that an spectroscopic UV feature of iron, the Fe III$\lambda\lambda$2039-2113 emission line blend, appears redshifted by typical amounts of $\sim 1000\rm\,km\,s^{-1}$ in many AGNs and quasars (Mediavilla et al. 2018, 2019, 2020). {Several evidences (see Appendix \ref{appendix}) from microlensing (Guerras et al. 2013, Fian et al. 2018), photoionization calculations (Temple et al. 2020),  modeling of the Fe III$\lambda\lambda$2039-2113 emission line profile (Mediavilla et al. 2018, 2019), and reverberation mapping (RM, Mediavilla et al. 2018) show that this blend originates close to the SMBH.} This UV iron blend allows to perform single object measurements of the gravitational redshift with very high significance: a mean value of 6.5$\sigma$ and a maximum of 15$\sigma$ for individual Sloan Digital Sky Survey (SDSS) quasar spectra with S/N$\gtrsim 20$\footnote{The possibility of measuring with good S/N ratio the gravitational redshift in single objects opens very interesting experimental opportunities (compare with the results of clusters of galaxies, in which after averaging huge numbers of galaxies the significance level of the gravitational redshift detection is not greater than 3$\sigma$ [Alam et al. 2017]).}.

{The test we propose and perform here is the iconic gravitational redshift experiment\footnote{Consequently, we test if the test-body trajectories are the geodesics of the metric (see, e.g., Misner, Thorne \& Wheeler 1973) or, more specifically, we test both pieces of EEP: the Weak Equivalence Principle and the local position invariance (Will 2014).} corrected to take into account the transverse Doppler effect\footnote{Transverse Doppler is a prediction of the special theory of relativity and, hence, its generalization is also a consequence of EEP (specifically of the local Lorentz invariance [Will 2014]).} arising from the orbital motion of the ionized gas around the SMBH. Let us write the measured gravitational plus transverse Doppler redshift as,

\begin{equation}
\label{redshiftm}
z^m_g=\left( {\Delta \lambda \over \lambda}\right)_ {Fe\, III}.
\end{equation}
From EEP\footnote{The velocity of the emitters needed to compute the transverse Doppler is taken from the Newtonian limit (Equation \ref{redshiftp} has been calculated for circular orbits [Mediavilla \& Insertis 1989] but the same result is obtained applying the virial theorem).} the predicted gravitational redshift is (Mediavilla \& Insertis 1989), 

\begin{equation}
\label{redshiftp}
z^p_g={3\over 2}{GM_{BH}\over c^2 R_{Fe\, III}}.
\end{equation}
Thus, to compare the observed and predicted gravitational redshifts we need independent estimates of the SMBH mass, $M_{BH}$, and of the size of the region emitting the Fe III blend, $R_{Fe\, III}$. There are several options to estimate SMBH masses (Mej\'\i a-Restrepo et al. 2018, Campitiello et al. 2020): to apply the virial theorem from the Doppler broadening of the emission lines, to use the $M_{BH}-\sigma_*$ or $M_{BH}-M_{bulge}$ relationships, or to model the quasar emission, for instance. Here, we consider the {first option (virial mass estimate)}. In this case, for the {emission line of} species $\chi$,
the BH mass is given by,

\begin{equation}
\label{virial}
M_{BH}=f_\chi {FWHM^2_\chi R_\chi\over G},
\end{equation}
where $FWHM_\chi$ and $R_\chi$ are the full width at half maximum of the spectral line used to estimate the mass, and the radial distance from the SMBH of the emitting region, respectively.  $f_\chi$ is the virial factor, which accounts for the geometry and kinematics of the emitting region. Thus\footnote{Alternatively, we may start, following the Newtonian version of the Weak Equivalence Principle, from the virial equation (Eq. \ref{virial}), modified to take into account possible differences between the inertial, $m$, and gravitational, $m_g$, masses, $M_{BH}=({m/m_g})f_\chi {FWHM^2_\chi R_\chi\over G}$, and use Eqs. \ref{redshiftm} and \ref{redshiftp} to estimate $M_{BH}$ to, finally, obtain ${m/m_g}={2\over 3} f^{-1}_\chi {R_{Fe\, III} \over R_\chi}{\left( {\Delta \lambda \over \lambda}\right)_ {Fe\, III} \over (FWHM_\chi/c)^2}$.},

\begin{equation}
\label{ratio}
{z^m_g\over z_g^p}={2\over 3} f^{-1}_\chi {R_{Fe\, III} \over R_\chi}{\left( {\Delta \lambda \over \lambda}\right)_ {Fe\, III} \over (FWHM_\chi/c)^2},
\end{equation}
is the predicted ratio to be experimentally tested.}

The paper is organized as follows. In \S \ref{results} we determine $z^m_g/ z_g^p$ (Eq. \ref{ratio}) using quasar spectra from X-shooter@VLT, SDSS and BOSS databases and UV spectra from two AGN: NGC 5548 and NGC 7469. In \S \ref{discussion} we compare our estimates with previous results at low cosmological redshift and discuss the cosmic evolution of EEP. Finally, in \S \ref{conclusions} we summarize the main conclusions.

\section{Results\label{results}}

%

According to Equation \ref{ratio}, we can estimate the ratio between the observed and predicted gravitational redshifts using widths corresponding to the Fe III$\lambda\lambda$2038-2113 blend or to other emission lines. In the latter case, both, the value of the virial factor, $f_\chi$, and the ratio between the emitting regions, ${R_{Fe\, III}/R_\chi}$, are needed. If we use, instead, Fe III$\lambda\lambda$2038-2113 data to estimate the widths, we only need to know the virial factor, $f_{Fe\, III}$. However, we lack independent measurements of $f_{Fe\, III}$ and we need a procedure to derive it.

\subsection{Results for  X-shooter@VLT quasar spectra (Capellupo et al. 2015, 2016) using the MgII, H$\beta$ and H$\alpha$ emission lines to estimate the FWHM\label{linesa}}

{It is convenient to start applying Eq. \ref{ratio} to the available dataset with the widest wavelength coverage and with a high S/N ratio: a sample of 10 quasars at $z\sim 1.55$ observed with X-shooter@VLT  (Capellupo et al. 2015, Capellupo et al. 2016) for which the redshifts of the Fe III$\lambda\lambda$2039-2113 blend have been measured (Mediavilla et al. 2019). 
Mg II, H$\beta$ and H$\alpha$ widths are also available (Mej\'\i a-Restrepo et al. 2016). To estimate the ${R_{Fe\, III} /R_{H\beta}}$ ratio we adopt {the commonly used} radius-luminosity relationship (Bentz et al. 2013), which scales H$\beta$ sizes with luminosities at $\lambda5100$,

\begin{equation}
\label{Bentz}
R_{H\beta}=391.74^{+28.99}_{-26.99}\left({\lambda L_{\lambda 5100}\over 10^{46}\rm\, erg\, s^{-1}}\right)^{0.533}\rm light-days.
\end{equation}
For the Fe III$\lambda\lambda$2039-2113, we adopt a similar scaling relationship anchored to the probability density function of the microlensing based size estimates obtained for a sample of lensed quasars by Fian et al. (2018),  

\begin{equation}
\label{rfeiii}
R_{Fe\, III}=17.35^{+8.39}_{-6.73}\left({\lambda L_{\lambda 1350}\over 10^{45.79}\rm\, erg\, s^{-1}}\right)^{0.533}\rm light-days,
\end{equation}
where 17.35 light-days and $10^{45.79}\rm\, erg\, s^{-1}$ are, respectively, the average of the Fe III sizes and $\lambda L_{\lambda 1350}$ luminosities of the quasars considered by Fian et al. (2018). The use for the Fe III case of the same exponent than for H$\beta$  is supported by previous studies (Mediavilla et al. 2018), which find a best-fit value of $0.57\pm 0.08$. Using the average values for $L_{\lambda 5100}$ and $L_{\lambda 1350}$ of the sample of 10 quasars observed with X-shooter, we obtain an average value of the ratio between sizes, $\langle R_{Fe\, III}/R_{H\beta}\rangle=0.106$, remarkably coincident with the photoionization model prediction, which indicates that the Fe III emission originates 1 dex closer to the SMBH than the CIV one\footnote{Notice, however, that any comparison with the CIV reverberation lags should be taken with caution as this line is kinematically complex and may present a peculiar response to variability owing to inclination effects.} (Temple et al. 2020). {To obtain the $\langle R_{Fe\, III}/R_\chi\rangle$ ratio for the other emission lines} we take $R_{H\alpha}=R_{CIV}=R_{Mg II}/2=R_{H\beta}$ (from RM estimates, Shen et al. 2019)\footnote{To support this equivalence between sizes, notice that according to Eq. \ref{virial} (adopting the same virial factor) sizes of different species should scale with the squared widths and if we apply this condition to the  X-shooter quasars sample, we obtain $R_{MgII}=2.1R_{H\beta}$.}.
Finally, we need an estimate of the virial factor, for which we take $\langle f \rangle=0.93\pm0.09$ {(see Appendix \ref{appendixB})}, obtained {averaging several values from the literature (Mej\'\i a-Restrepo et al. 2018, Collin et al. 2006, Graham et al. 2011, Ho \& Kim 2014, Woo et al. 2015, Williams et al. 2018, Wang et al. 2019, Yu et al. 2020)}.  The resulting average ratio between the observed and predicted gravitational redshift for the X-shooter quasars sample, $\langle {z^m_g/z_g^p}\rangle=0.87\pm0.12$, at $z=1.55$ is shown (light red open circle) in Figure \ref{fig} (see also Table \ref{tableratios}).
}

\subsection{Results for SDSS quasar spectra using the MgII emission lines to estimate the FWHM\label{linesb}}

{In a second step, we apply Eq. \ref{ratio} to a sample of quasars from the SDSS, the main available source of quasar spectra. We select the 85 spectra that fulfill the following criteria: S/N$>$20,  clear detection of the Fe III$\lambda\lambda$2039-2113 blend (S/N$>$3 at the peak of the feature) and presence of the Mg II line in the observed wavelength range. Following the procedure described in Mediavilla et al. (2018), we fit the Fe III$\lambda\lambda$2039-2113 blend to these spectra,  determining redshifts of this iron feature in the range $\Delta\lambda_{Fe III}=+2.2\,$\AA\ to $+15\,$\AA, with median $+6.6\,$\AA. We also fit the Mg II emission line according to the procedure detailed in Mediavilla et al. (2019).
 We apply Eq. \ref{ratio} using the averaged size ratio, $\langle {R_{Fe\, III} /R_{Mg II}}\rangle$, and virial factor,  $\langle f \rangle$, derived in \S \ref{linesa}. 
In Figure \ref{fig} (see also Table \ref{tableratios}) we present (light red squares) averages of $z^m_g/z_g^p$ at 6 different cosmological redshift bins of the SDSS single spectra data. 
%
}


\subsection{Results for BOSS composite quasar spectra using the C IV and Mg II  emission lines to estimate the FWHM\label{linesc}}

{To go further in cosmological redshift we use the gravitational redshifts of the Fe III$\lambda\lambda$2039-2113 blend of the high S/N baryon oscillation spectroscopic survey (BOSS) composite spectra (Jensen et al. 2016) estimated {in Mediavilla et al. (2018)}. We have used the widths of both emission lines, CIV (Jensen et al. 2016) and Mg II (Mediavilla et al. 2019). The ${z^m_g/z_g^p}$ ratios averaged in 3 cosmological redshift bins are shown in Figure \ref{fig} (light red triangles) and in Table \ref{tableratios}.}


\subsection{Results for NGC 5548 \label{5548} and NGC 7469 using H$\beta$ emission lines to estimate the FWHM}

{At low cosmological redshift, we have estimates (based on UV observations) of the Fe III$\lambda\lambda$2039-2113 gravitational redshifts for the nearby AGN NGC 5548 and NGC 7469 (Mediavilla et al. 2018). Taking average  values of the $\rm FWHM_{H\beta}$ ($6612\pm 1646 \rm\, km/s$ for NGC 5548 and $4267\pm 719\, \rm km/s$ for NGC 7469, Yu et al. 2020), we obtain ${z^m_g/ z_g^p}=0.88\pm 0.35$ for NGC 5548 and $0.98\pm0.30$ for NGC 7469 (see red open hexagons in Figure \ref{fig}). Notice that the uncertainties of these single object measurements, are comparable to those obtained from the statistics of cluster galaxies. }

{According to Li et al. (2016) and Bon et al. (2016), NGC 5548 could host a supermassive BH binary and our redshift determination in NGC 5548 may be affected by its orbital motion. Notice, however, that monitoring of the H$\beta$ line shows that even if the centroid can shift between epochs its broad component (specifically the full width at 25\% of the maximum intensity) appears systematically redshifted  (Bon et al. 2016). This result limits the impact of the possible orbital motion on the gravitational redshift estimates.}

\subsection{Results for SDSS and BOSS quasar spectra using the Fe III$\lambda\lambda$2038-2113 line emission blend to estimate the FWHM}

{Finally, we can also calculate the ${z^m_g/ z_g^p}$ ratio using the widths of the Fe III UV blend itself in Eq.\ref{ratio}.   However, as the Fe III$\lambda\lambda$2039-2113 seems to arise from a region (likely the accretion disc) very  compact and with different geometry than the region emitting the broad emission lines, we can not extrapolate the information available in the literature about the virial factor of this last region. Fortunately, in the SDSS  data we find a rather good correlation between the FWHMs of Fe III$\lambda\lambda$2039-2113 and of Mg II (which confirms that the Fe III$\lambda\lambda$2039-2113 is a virial indicator, as the Mg II is). Then, applying the virial equation (Eq. \ref{virial}) to both species, averaging over the widths and equalizing the resulting masses, we derive, $f_{FeIII}=(R_{MgII}/R_{FeIII})(\langle FWHM^2_{FeIII}\rangle/\langle FWHM^2_{MgII}\rangle)f_{MgII}$. Inserting in this equation the averaged size ratio, $\langle {R_{Fe\, III} /R_{Mg II}}\rangle$, and virial factor,  $\langle f \rangle$, derived in \S \ref{linesa}, we estimate $f_{Fe III}=17.6$ {(adopting a geometrical interpretation, this high value of the virial factor would correspond to a flattened structure oriented almost face-on, $\langle i \rangle \sim 14^{o}$, supporting the origin of the Fe III$\lambda\lambda 2039-2113$ blend in the accretion disk)}. Then, we apply Eq. \ref{ratio} to the ($\Delta \lambda_ {Fe\, III}$, $FWHM_{Fe III}$) pairs determined from the SDSS and BOSS spectra. The resulting ratios, averaged in 6 redshift bins for the SDSS data and in 3 for the BOSS data, are shown in Figure \ref{fig} (light magenta diamonds) and Table \ref{tableratios}. Obviously, all these data points are linked to the Mg II  results by a global factor, through the derivation of $f_{Fe III}$,  but the relative differences between bins are not, i.e., the ratios inferred from the Fe III$\lambda\lambda$2039-2113 FWHMs can independently be used to test the cosmic evolution of EEP.}

\section{Discussion\label{discussion}}

{
Our experimental results to test EEP along with others from the literature are summarized in Figure \ref{fig}, in which two remarkable results can be readily seen: (i) the wide coverage in cosmological redshift of our data compared with the tiny region around $z_{cosm}\sim 0.22$ previously studied and (ii) the comparatively good statistical uncertainties of our measurements (see also Table \ref{tableratios}), which range from $\sim9\%$ to $\sim20\%$ ($\sim$35\% if we include the NGC 5548 and NGC 7469 data at $z_{cosm}\sim 0.017$), similar or better than the errors of other measurements outside the Solar System.


{Regarding EEP, we do not observe any statistically significant trend with cosmological redshift.} Averaging the data corresponding to C IV, Mg II, H$\beta$ and H$\alpha$ in the  $z_{cosm}\sim 1$ to 3 redshift range, we obtain a mean value $\langle z^m_g/z_g^p\rangle=1.02\pm 0.07$ with a scatter of the data of $0.21\pm 0.05$.
We obtain similar results averaging the Fe III$\lambda\lambda$2039-2113 data,  $\langle z^m_g/z_g^p\rangle=1.06\pm 0.08$ with a scatter of $0.25\pm 0.06$. The average value corresponding to Fe III$\lambda\lambda$2039-2113 is tied to Mg II through the derivation of $f_{Fe III}$, but the comparison between the ratios at different cosmological redshifts inferred from the Fe III$\lambda\lambda$2039-2113 widths is an independent confirmation
of the constancy of EEP along the studied cosmological redshift range. 

So far we have considered the data grouped in 10 bins in $z_{cosm}$ respecting the data source (see Figure \ref{fig} and Table \ref{tableratios}).  It is convenient to increase the S/N ratio by regrouping the data, irrespective of their origin, in 4 $z_{cosm}$ bins, shown as red points in Figure \ref{fig} (see also Table \ref{tableratios2}).
  As expected, the mean does not change significantly with the regrouping, $\langle z^m_g/z_g^p\rangle=1.05\pm 0.06$, while the scatter improves by almost a factor 2 ($0.13\pm 0.05$). Thus, we find that the gravitational redshift derived from EEP holds with less than a $13\%$ of deviation in this cosmological redshift range, and, averaging the whole redshift range (blue data point in Figure \ref{fig}), we validate EEP with an statistical uncertainty of 6\%.



Although until now we have only taken into account statistical (random) errors, there are also systematic uncertainties in our estimates arising from the factor  ${R_{Fe\, III} / (f_\chi R_\chi)}$ in Equation \ref{ratio}. This factor does not affect at all to the study of the cosmic evolution as we can remove it by normalizing the  $z^m_g/z_g^p$ ratios to the mean\footnote{A sort of variant, between epochs, of the "null redshift experiment".}:  $(z^m_g/z_g^p)/\langle z^m_g/z_g^p\rangle$ . With respect to the averaged values, $\langle z^m_g/z_g^p\rangle$, we could anchor them to $z\sim 0$ where we know from other redshift experiments that EEP holds\footnote{Using, for instance, the average of our measurements of NGC 5548 and NGC 7469 as baseline to do the anchoring we would obtain $\langle z^m_g/z_g^p\rangle=1.13\pm 0.09$ for the mean in the $z_{cosm}=1-3$ range.}.
In any event, the evaluation and mitigation of systematics errors is interesting if one wants to consider each measurement separately from the others. The budget of systematic uncertainties is largely dominated by the error in $R_{Fe \ III}$, roughly inferred from microlensing measurements (from Eq. \ref{rfeiii} we estimate an uncertainty of $\sim 44\%$) but  that can be greatly reduced (also the uncertainty in $R_\chi$ and $f_\chi$) by precise RM measurements or complementary observations. Moreover, the factor  $R_\chi f_\chi$ can be eliminated by using a different method to estimate BH masses in Eq. \ref{redshiftp} (e.g., using $M_{BH}-\sigma_*$ or $M_{BH}-M_{bulge}$ relationships, or quasar emission modeling\footnote{Although in this case, obviously, other systematic errors may appear instead.}).

On the other  hand, the statistical uncertainties can also be greatly reduced with the help of new observations (notice that we are just considering about 10 spectra per redshift bin!). Most of the used data come from the SDSS and, hence, have been obtained with a modest size telescope, but observation of $\sim1000$ sources with a $4-10\,\rm m$ class telescope could bring statistical uncertainties down to the 1\% level. }

\section{Conclusions\label{conclusions}}


{We propose a new method to test EEP in a large range in cosmological redshift. We explore an uncharted cosmic territory covering $\sim 4\times 10^9$ years of the history of the universe (in principle, the method can be extended to other epochs  using quasar spectra with an adequate wavelength coverage). We show the feasibility of the method, obtaining errors comparable or better than other classical methods applied outside the SS. The main results are:

1. We measure the gravitational redshift directly in individual objects (85 SDSS quasars among them) in the early universe (as far as 10 Gyr ago). These single object measurements have a high statistical significance ($>> 3\sigma$). 

2. At low redshift, we estimate
the ratio between the observed gravitational redshifts and the theoretical predictions in the AGN NGC 5548 and NGC 7469 ($z^m_g/z_g^p=0.88\pm 0.35$ and $z^m_g/z_g^p=0.98\pm 0.30$, respectively). These measurements, on their own, represent a qualitative leap to test EEP.

3. For the first time we measure the ratio between the observed gravitational redshifts, and the theoretical predictions, $z^m_g/z_g^p$, in 10 independent cosmological redshift bins covering the $z_{cosm}\sim 1$ to 3 range. 

4. We find that EEP predictions for the gravitational redshift hold in this cosmological redshift range ($1\lesssim z_{cosm} \lesssim 3$) with an statistical uncertainty of 6\%, without evidence for cosmic evolution above 13\%.



The present results can be extended to fill the gap at cosmological redshifts below 1 with UV observations. On the other edge, the gravitational redshift of massive quasars detected at high cosmological redshift (up to $z_{cosm}\sim 7$), could be probed with IR observations. In the future, application of our method to larger datasets and better quality spectra will open extraordinary possibilities for precise tests of EEP over most of the history of the universe}

\acknowledgements{{We thank the referee for the thorough and constructive review of the paper.} We thank SDSS and BOSS surveys for providing data. We tank Temple et al. for providing the data of their narrow and broad composites. This research was supported by the Spanish MINECO with the grants AYA2016-79104-C3-1-P and AYA2017-84897-P, by the Fondo Europeo de Desarrollo Regional (FEDER), and by project FQM-108 financed by Junta de Andalucia.}

\appendix
\section{Origin of the Fe III$\lambda\lambda$2039-2113 emission line blend. \label{appendix}} 

{There are several evidences that the Fe III$\lambda\lambda$2039-2113 blend originates close to the SMBH:  (i) it shows very strong gravitational microlensing (Guerras et al. 2013, Fian et al. 2018) with magnifications comparable to those of the continuum generated by the accretion disk; (ii) according to photoionization calculations (Temple et al. 2020), the UV emission of the Fe III ion traces especially high density gas confined in a region very close to the central SMBH and moving in quasi-ordered flows in the equatorial plane of the AGN (likely in the accretion disk, Temple et al. 2020); (iii) the profile of the Fe III$\lambda\lambda$2039-2113 blend is very well fitted with a single kinematic component of the standard template of the blend (Vestergaard \& Wilkes 2001), supporting the confinement of the gas and explaining the clean measurement of the gravitational redshift (see below the fit to the SDSS narrow and broad composite spectra constructed by Temple et al. 2020), (iv) the observed redshifts match not only the expected magnitude of the gravitational redshift but also the theoretical correlation between the gravitational redshift and the squared widths of the emission lines\footnote{In the case of the Fe II blend an artificial correlation (arising from a combination of noise and inadequate modeling of the stellar host in the spectra) between line shifts and widths has been reported by  Bon et al. (2020). However, in the case of the Fe III$\lambda\lambda2039-2113$ blend we find that the redshifts correlate not only with the widths of the Fe III$\lambda\lambda 2039-2113$ itself, but also with the widths of the emission lines of other species, like H$\alpha$, H$\beta$ or MgII, which strongly supports the physical origin of the observed correlations.} (Mediavilla et al. 2018, 2019) (i.e. correlate with $M_{BH}/R$); and (v) the only reverberation mapping estimate available (Mediavilla et al. 2018) of the size of the region emitting the Fe III$\lambda\lambda$2039-2113 blend, confirms that this region has a size comparable to that of the continuum. 

The narrow and broad high S/N composites created by Temple et al. (2020) by averaging SDSS spectra with $FWHM_{Mg II}$ in the range $2000-4000\,\rm km\,s^{-1}$ (narrow) and with $FWHM_{Mg II}$ in the range $9000-15000\,\rm km\,s^{-1}$ (broad) are perfectly suited to illustrate several results of the study of the Fe III$\lambda\lambda$2039-2113 blend: the good matching of the line profile with a single kinematic component and the existence of a gravitational redshift (larger for wider lines). 
%
%
 We model the blend in both composite spectra  (kindly provided by Temple et al. 2020) using the standard template (Vestergaard \& Wilkes 2001) and following the same steps as in Mediavilla et al. (2018). The resulting fits (notice the remarkable good matching, specially of the characteristic features of the narrow composite shape) are presented in Figure \ref{ferlandcomposites}. Both composites are redshifted with respect to the systemic velocity of the quasar: the narrow by $+3.95\,$\AA\ and the broad by a larger, as expected, quantity, $+7.54\,$\AA. We measure broadenings of $2163\,\rm km\,s^{-1}$ and $4632\,\rm km\,s^{-1}$. 

In Figure \ref{ferlandcomparison} we compare the original broad composite blend (blue points) with the narrow blend convolved with a Gaussian to match the broadening of the broad composite according to the widths estimated from the fits (orange dotted line). The redshift of the blue points respect to the orange dotted line is evident. The green dotted line is the orange dotted line shifted by the difference in redshifts estimated from the fits. It matches fairly well the blue points.

{Finally, we would like to remark the striking good fits based on the template of Vestergaard \& Wilkes (2001) to a variety of objects with quite different properties (masses, emission line broadenings, degree of activity, etc.).  The Vestergaard \& Wilkes (2001) template is inferred from IZw1, a narrow line Seyfert 1 galaxy, which may be not representative of objects with very broad emission lines. However, we have applied this template to many AGNs and quasars (cfr. \S \ref{results}, see also Mediavilla et al. 2018, 2019, 2020) obtaining surprisingly good fits if we take into account the scatter in properties among these objects. Moreover, in Mediavilla et al. (2018) we obtain consistent fits to other UV Fe III features like the Fe III$\lambda\lambda1970-2039$ blend and the Fe III$\lambda2419$ emission line. These results seem to indicate that the UV Fe III emission arises from a homogeneous region with quite regular physical conditions.}

{\section{Estimate of the average virial factor. \label{appendixB}} 

The used value of the virial factor, $f$, has been calculated as the weighted mean of the estimates given in the following references (see Figure \ref{figure_virial}): Mej\'\i a-Restrepo et al. (2018), Collin et al. (2006), Graham et al. (2011), Ho \& Kim (2014), Woo et al. (2015), Williams et al. (2018), Wang et al. (2019) and Yu et al. (2020).  (Notice that we use two values from Collin et al. (2006), one of them based in the dependence with the FWHM). Each estimate is weighted according to $1/\sigma_f^2$, where $\sigma_f$ is the error estimate given in the original data source. (For Mej\'\i a-Restrepo et al. [2018] and Collin et al. [2006] we use the $f$ versus FWHM relationships given by these authors to estimate $f$ and $\sigma_f$ using the average FWHM derived from our X-shooter sample). The resulting weighted mean is $\langle f \rangle=0.93\pm 0.09$ with scatter $0.25\pm 0.06$. Using a non-weighted average instead produces a similar value, $\langle f \rangle=0.99\pm 0.08$. }

\clearpage

\begin{figure}[h]
\includegraphics[scale=1.1]{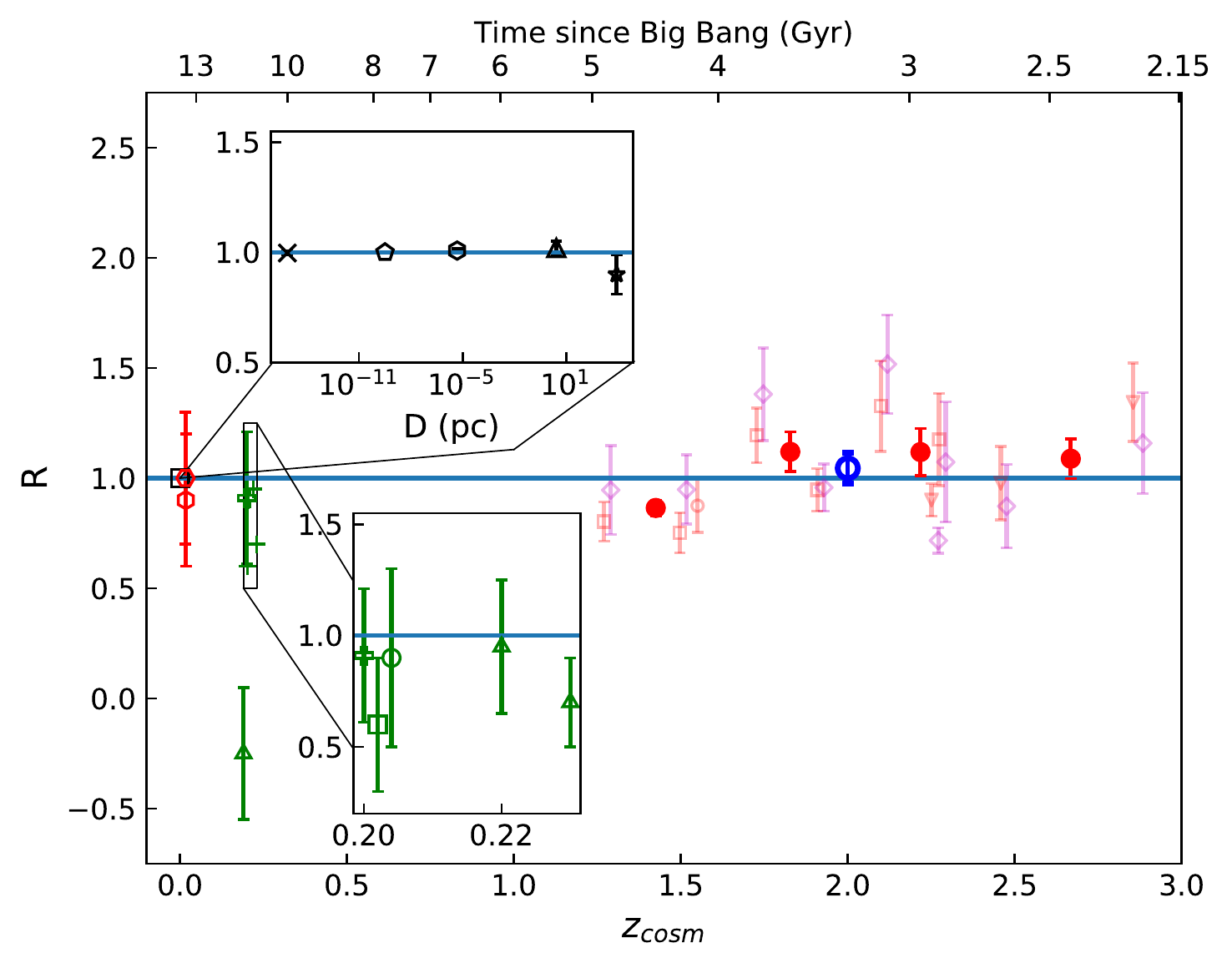}
\caption{Ratio between the measured and predicted gravitational redshift, $R=z^m_g/z_g^p$, vs. cosmological redshift, $z_{cosm}$. {\bf Main panel} (data from the present work; see text and Tables \ref{tableratios} and \ref{tableratios2}): NGC 5548 and NGC 7469 ({\color{red} $\hexagon$}), SDSS quasars ({\color{red} $\square$}), BOSS composite quasar spectra ({\color{red} $\nabla$}), X-shooter quasars ({\color{red} $\bigcirc$}), SDSS and BOSS estimates based on Fe III widths ({\color{magenta} $\Diamond$}), averages in 4 redshift bins ({\color{red} $\bullet$}), average in all the $z_{cosm}=1-3 $ range ({\color{blue} $\bigcirc$}). {\bf Superior inset} (Solar System and Milky Way data from the literature; $z_{cosm}=0$): ground experiment ($\times$, Pound \& Snider 1964),  space-borne experiment (\pentagon, Vessot et al. 1980), Sun (\hexagon, Gonz\'alez Hern\'andez 2020), Sirius B ($\triangle$, Joyce et al. 2018), S2-Galactic centre ($\largestar$, Gravity collaboration et al. 2018). {\bf Inferior inset} (galaxy cluster data from the literature; $z_{cosm}\sim 0.22$): {\color{green} $\triangle$} (Jimeno et al. 2015),  {\color{green} $+$} (Wojtak et al. 2011), {\color{green} $\square$}  (Kaiser 2013), {\color{green} $\bigcirc$} (Sadeh et al.  2015). \label{fig}}
\end{figure}

\clearpage

\begin{figure}[h]
\includegraphics[scale=1.0]{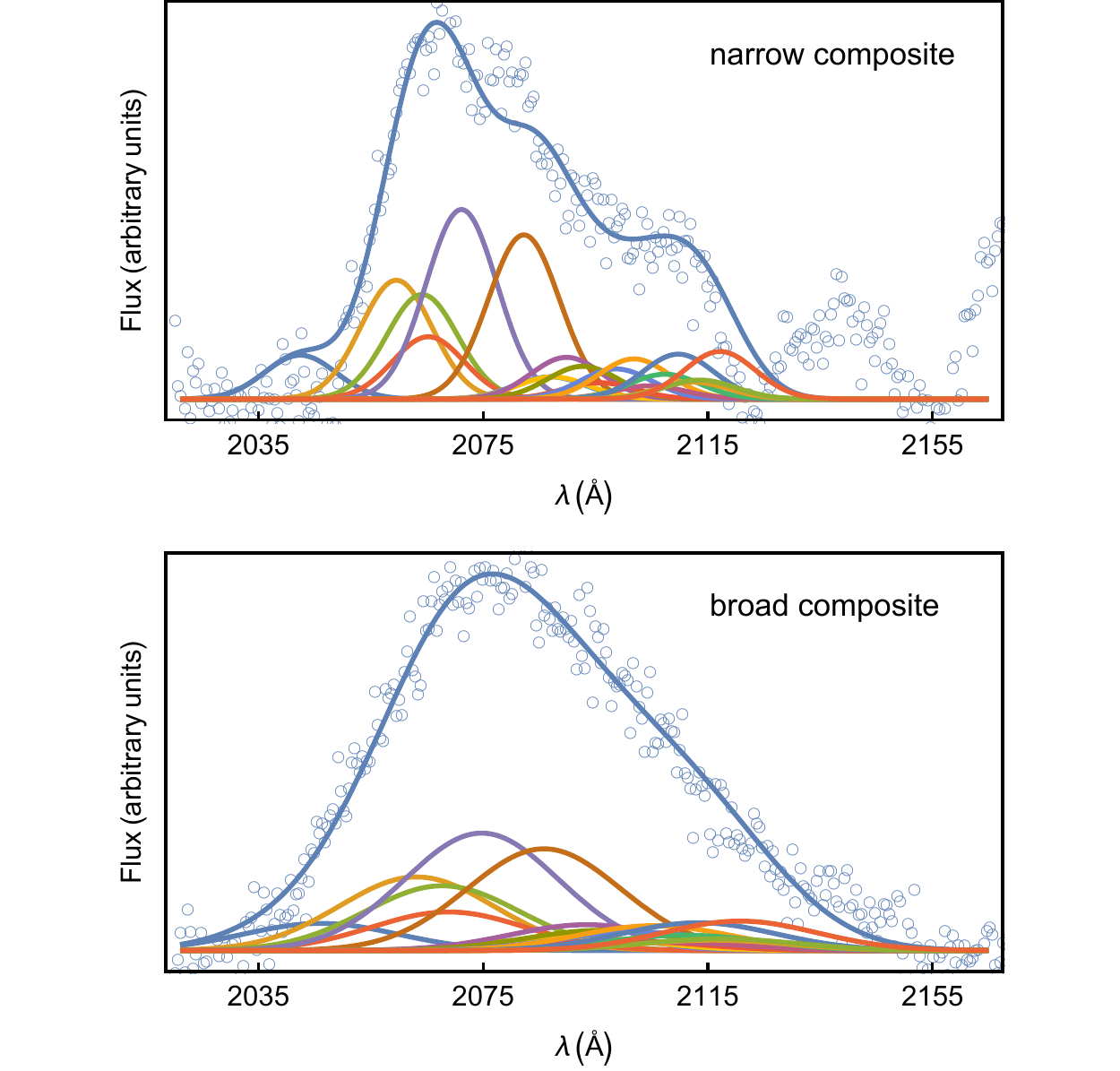}
\caption{Fits of the Vestergaard \& Wilkes (2001) template of the Fe III$\lambda\lambda$2039-2113 blend to the narrow and broad composite spectra constructed by Temple et al. (2020) from SDSS spectra (see Appendix \ref{appendix}). Open circles correspond to the data, the blue line to the
template and the other lines to the Gaussians representing each of the Fe III lines. \label{ferlandcomposites}}
\end{figure}

\clearpage

\begin{figure}[h]
\includegraphics[scale=1.1]{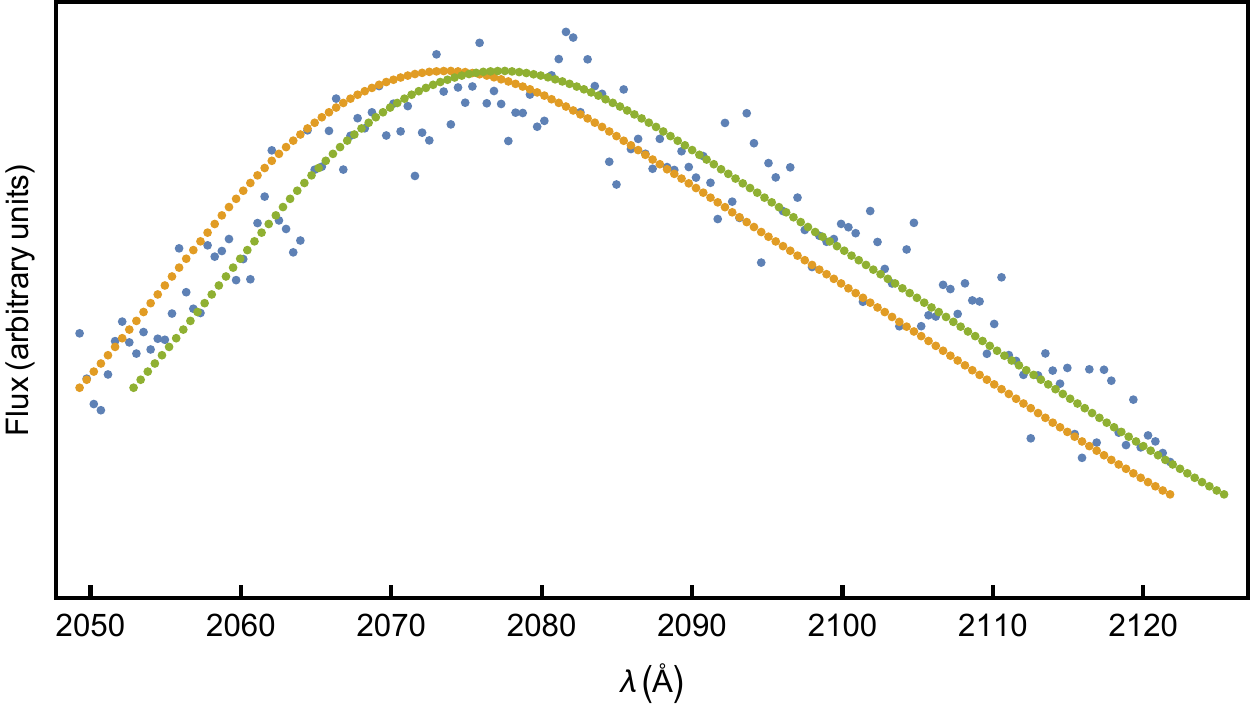}
\caption{Comparison, in the spectral region of the Fe III$\lambda\lambda$2039-2113 blend, between the broad composite of Temple et al. (2020) (blue points) and the narrow composite convolved with a Gaussian (orange dotted line) to match the broadening of the broad composite according to the fits (see Figure \ref{ferlandcomposites}; see Appendix \ref{appendix}). The green dotted line is the orange dotted line redshifted to match the redshift of the broad composite according to the fits (see Figure \ref{ferlandcomposites}; see Appendix \ref{appendix}). \label{ferlandcomparison}}
\end{figure}

\clearpage

\begin{figure}[h]
\includegraphics[scale=1.1]{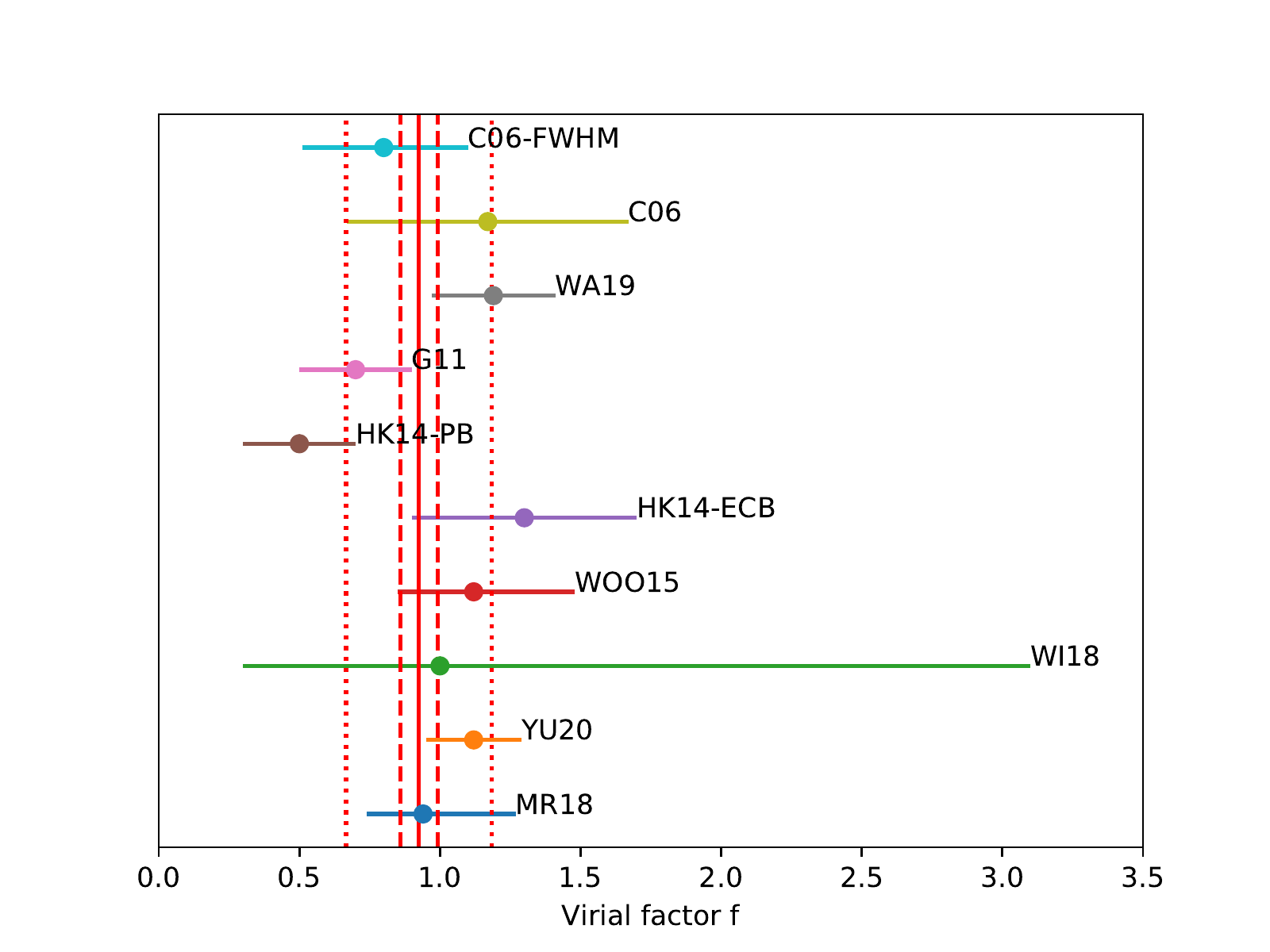}
\caption{Individual estimates of the virial factor for different sources (points with $\pm 1\sigma$ error bars), and weighted average (vertical red line). The dashed vertical red lines correspond to $\pm1$ standard error of the mean and the dotted ones to $\pm 1$ standard deviation (see Appendix \ref{appendixB}). \label{figure_virial}}
\end{figure}

\clearpage

\begin{table}
\centering
\caption{Measured vs. predicted gravitational redshift ratios for 10 $z_{cosm}$ bins}
\medskip
\begin{tabular}{ccccc}
\hline
$\langle z_{cosm}\rangle$ & $\langle{z^m_g/z_g^p}\ (CIV,MgII,H\beta,H\alpha)\rangle$ &$\langle z_{cosm}\rangle$& $\langle {z^m_g/z_g^p}\ (FeIII)\rangle$& Data Source \\
\hline
$1.27$& $0.80\pm0.09$ &$1.29$& $0.95\pm0.21$ &SDSS \\
$1.50$ & $0.75\pm0.09$ &$1.52$& $0.95\pm0.16$& SDSS \\
$1.55$ & $0.87\pm0.12$ &$--$&  $---$ &X-shooter \\
$1.73$& $1.19\pm0.12$ &$1.75$& $1.38\pm0.22$ &SDSS \\
$1.91 $& $0.94\pm0.10$ &$1.93$& $0.96\pm0.11$& SDSS \\
$2.10$ & $1.32\pm0.21$ &$2.12$&  $1.52\pm0.23$ &SDSS \\
$2.25$& $0.89\pm0.07$ &$2.27$& $0.72\pm0.06$ &BOSS \\
$2.27 $& $1.17\pm0.21$ &$2.29$& $1.07\pm0.28$& SDSS \\
$2.46 $& $0.97\pm0.17$ &$2.48$&  $0.87\pm0.20$ &BOSS\\
$2.85$& $1.34\pm0.18$ &$2.88$& $1.16\pm0.24$ &BOSS \\

\hline
\end{tabular}
\label{tableratios}
\end{table}

\begin{table}
\centering
\caption{Measured vs. predicted gravitational redshift ratios for 4 $z_{cosm}$ bins}
\medskip
\begin{tabular}{cc}
\hline
$\langle z_{cosm}\rangle$ & $\langle{z^m_g/z_g^p}\ (CIV,FeIII,MgII,H\beta,H\alpha)\rangle$  \\
\hline
$1.42$& $0.86\pm0.03$  \\
$1.83$ & $1.12\pm0.09$  \\
$2.22$ & $1.12\pm0.11$ \\
$2.67$& $1.09\pm0.09$  \\
\hline
\end{tabular}
\label{tableratios2}
\end{table}


\end{document}